\documentclass[11pt]{revtex4}

\newcommand{\N}{N\raise.7ex\hbox{\underline{$\circ $}}$\;$}

\def\ch{\cosh}
\def\sh{\sinh}

\begin{document}

\title{
  Parameters
    of Lorentz Matrices and Transitivity  in Polarization Optics}

\author{E.M. Ovsiyuk}
\author{O.V. Veko}
\author{M. Neagu}
\author{V. Balan}
\author{V.M. Red'kov}




\begin{abstract}

In the context of applying the Lorentz group theory to
polarization optics in the
 frames of Stokes--Mueller formalism, some properties of the Lorentz group are investigated.
We start with the factorized form of arbitrary Lorentz matrix as a
product of two  commuting and  conjugate $4\times 4$-matrices,
$L(q,q)=A(q_{a})A^*(q_{a});\; a= 0,1,2,3$. Mueller matrices of the
Lorentzian type $M=L$ are pointed out as a special sub-class in
the total set of $4\times 4$ matrices of the linear group
$GL(4,R)$. Any arbitrary Lorentz matrix is presented as a linear
combination of 16 elements of the Dirac basis. On this ground, a
method to construct parameters $q_a$  by an explicitly given
Lorentz matrix L is elaborated.
It is shown that the factorized form of L=M  matrices
 provides us with a number of simple transitivity equations relating couples of initial and final
 4-vectors, which are defined in terms of parameters $q_a$  of the Lorentz group. Some of these
  transitivity relations can be interpreted within polarization optics and can be applied to the
   group-theoretic analysis of the problem of measuring Mueller matrices in optical experiments.

\end{abstract}

Report to:

The International Conference
 "Differential Geometry and Dynamical Systems - 2012".

 29 August - 2 September 2012, Mangalia, Romania

\maketitle

\section{On establishing the parameters of Lorentz matrices
from their explicit form}
In the context of applying the Lorentz group theory to polarization optics in the frames of
    Stokes--Mueller formalism, some properties of the Lorentz group are investigated (see
    \cite{1}-\cite{10};  the notation according to \cite{9,10} is used).
    We start with a factorized representation for Lorentz matrices\footnote{Hereafter,
    we denote by "*" the  complex conjugation.}
    \begin{eqnarray}[\; L_{b}^{\;\;a}(q,\bar{q}^*) \; ]  =A(q)  A^*(q)\;,\label{1.1}\end{eqnarray}
where
    \begin{eqnarray}A (q)=\left(\begin{array}{rrrr}
    q_0&-q_1&-q_2&-q_3\\
    -q_1& q_0&-iq_3&iq_2\\
    -q_2&iq_3& q_0&-iq_1\\
    -q_3&-iq_2&iq_1& q_0\end{array}  \right), \qquad
    A^*(q)  =\left (\begin{array}{rrrr}
    q_0^*&-q_1^*&-q_2^*&-q_3^*\\
    -q_1^*& q_0^*&iq_3^*&-iq_2^*\\
    -q_2^*&-iq_3^*& q_0^*&iq_1^*\\
    -q_3^*&iq_2^*&-iq_1^*& q_0^*\end{array}  \right).\label{1.2}\end{eqnarray}
We shall prove that one can easily construct matrices for which the complex vectors  $q$
    and  $q^*$ are eigenvectors. Indeed, the following identities hold\footnote{We denote $q=(q_0,\mathbf{q})
    \equiv(q_0,q_1,q_2,q_3)$, and recall that we further assume that $q_0^2-{\bf q}^2=+1$,
    where ${\bf q}^2=\langle{\bf q},{\bf q}\rangle$.}
    \begin{eqnarray}L \; q=\bar{q}^*\;, \qquad L \; q^*=\bar{q}\;,\qquad
    \bar{q}^*=( q_0^*, -{\bf q}^*)\;, \qquad \qquad  \bar{q}=( q_0, -{\bf q} )\; .\label{1.4a}\end{eqnarray}
With the use of the special element
    $$\delta=\left(\begin{array}{cccc}1&0&0&0\\0&-1&0&0\\0&0&-1&0\\0&0&0&-1\end{array} \right),$$
the relations from (\ref{1.4a}) read
    \begin{eqnarray}\delta \; L \; q=q^*\;, \qquad\delta \; L \; q^*=q\; .\label{1.4c}\end{eqnarray}
whence, through an elementary process we get
    $$\delta\; L \; \delta \; L \; q=\delta L\; q^*=q \;, \qquad
        \delta \; L\; \delta \; L \; q^*=\delta \; L\; q =q^* \; .$$
i.e.,
    \begin{eqnarray}Q=(\delta\; L )^2 \;, \qquad Q q=q\;, \qquad  Q q^*=q^* \; .
    \label{1.5b}\end{eqnarray}
Allowing the pseudo-orthogonality of Lorentz transformations\footnote{We further denote by dash the transposition
    of matrices.}, i.e., $\delta L \delta=(L^{t})^{-1}$,
    the matrix  $Q$ may be represented as
    \begin{equation}Q=(L^{t})^{-1} L \; .\label{1.7b}\end{equation}
In particular, for the orthogonal subgroup of rotations, we have
    \begin{equation}L=\left(\begin{array}{cc}1&0\\0&O\end{array} \right), \qquad
    (L^{t})=L^{-1}\;, \qquad Q=L^2= \left(\begin{array}{cc}1&0\\0&O^2\end{array} \right),
        \label{1.7c}\end{equation}
with $OO^t=I_3,\;\det O=1$. The explicit form for an arbitrary Euclidean rotation is fixed
    by the parameters $q_0=n_0, \; q_{j}=-i n_{j}$, so that
    $$L =\left(\begin{array}{rrrr}
    n_0&in_1&i n_2&i n_3\\i n_1&n_0&-n_3&n_2\\
    in_2&n_3&n_0&-n_1\\i n_3&-n_2&n_1&n_0\end{array} \right)
    \left(\begin{array}{rrrr}
    n_0&-in_1&-i n_2&-i n_3\\-in_1&n_0&-n_3&n_2\\
    -in_2&n_3&n_0&-n_1\\-i n_3&-n_2&n_1&n_0\end{array} \right)=
    \left(\begin{array}{cc} 1&0\\0&O\end{array} \right),$$
where
    \begin{equation}O =  \left(\begin{array}{lll}
    1 -2 (n_2^2+n_3^2)&-2n_0n_3+2n_1n_2&+2n_0n_2+2n_1n_3\\
   +2n_0n_3+2n_1n_2&1 -2 (n_3^2+n_1^2)&-2n_0n_1+2n_2n_3\\
    -2n_0n_2+2n_1n_3&+2n_0n_1+2n_2n_3&1 -2 (n_1^2+n_2^2)
    \end{array} \right) .\label{1.8b}\end{equation}
The  $4\times 4$-matrix $L$ acts on the 4-vector  $n_{a}$
according to
    \begin{equation}\left(\begin{array}{cc}1&0\\0&O\end{array} \right)
    \left(\begin{array}{r}n_0\\-i {\bf n}\end{array} \right)=
    \left( \begin{array}{r}n_0\\-i {\bf n}\end{array} \right).\label{1.8c}\end{equation}
Now let us consider arbitrary pseudo-Euclidean rotations -- these are fixed by the parameters
    $q_0=m_0 \;,\; q_{j}=m_{j}\;, \; m_0^2-{\bf m}^2=+1.$

By denoting ${\bf e}=(e_1,e_2,e_3)$, we have
    \begin{equation}m_0=\mbox{cosh}( \chi/2) \;, \qquad  m_{j}=\mbox{sinh}( \chi/2) \;e_{j},
        \qquad {\bf e}^2=1 \; ;\label{1.9a}\end{equation}
we get the following representation
    \begin{equation}L=\left(\begin{array}{cccc}
    \ch \chi&-\sh \chi e_1&-\sh \chi e_2&-\sh \chi e_3\\
    -\sh  \chi e_1&1+(\ch \chi -1) e_1^2&(\ch x-1)e_1e_2&(\ch x-1)e_1e_3\\
    -\sh \chi e_2&(\ch x-1)e_1e_2&1 +(\ch \chi -1) e_2^2&(\ch x-1)e_2e_3\\
    -\sh \chi e_3&(\ch x-1)e_3e_1&(\ch x-1)e_2e_3&1 +(\ch \chi -1) e_3^2
    \end{array} \right)  .\label{1.9a}\end{equation}
Then, we readily find by direct calculation that
    \begin{equation}Q=(\delta L)^2=  \left(\begin{array}{cccc}
    1&0&0&0\\0&1&0&0\\0&0&1&0\\0&0&0&1\end{array} \right) .\label{1.9c}\end{equation}
This identity can be proved in a different way. Since the pure Lorentz transformation
    is given by some symmetrical matrix, the relation (\ref{1.7b}) takes the form
\begin{eqnarray}Q=(L^{t})^{-1} L=L^{-1} L=I,\label{1.9d}\end{eqnarray}
which coincides with (\ref{1.9c}).

We shall further explicitly describe the case when the arbitrary Lorentz matrix is given by blocks
    $$
    L=\left(\begin{array}{cc}
    K&M\\N&L\end{array} \right),
    $$

\textbf{block} $(K)$ \begin{eqnarray}
    L_{00}=(q_0\,q_0^*+q_1\,q_1^*)+(q_2\,q_2^*+q_3\,q_3^*)\,,
        \qquad L_{01}=- (q_0\,q_1^*
       +q_1\,q_0^*)+i\,(q_2\,q_3^*-q_3\,q_2^*)\,,\nonumber\\
    L_{10}=-(q_0\,q_1^*+q_1\,q_0^*) -i\,(q_2\,q_3^*-\,q_3\,q_2^*) \,,
        \qquad L_{11}=(q_0\,q_0^*
   +q_1\,q_1^*)-(q_2\,q_2^*+q_3\,q_3^*)\,;\nonumber\label{1.12a}\end{eqnarray}
\textbf{block} $(M)$\begin{eqnarray}
    L_{22}=(q_0\,q_0^*-q_1\,q_1^*)+(q_2\,q_2^*
        -q_3\,q_3^*)\,,\qquad L_{23}=i\,(q_0\,q_1^*
        -q_1\,q_0^*)+(q_2\,q_3^*+q_3\,q_2^*)\,,\nonumber\\
    L_{32}=-i\,(q_0\,q_1^* -q_1\,q_0^*)+(q_2\,q_3^*
       +q_3\,q_2^*) \,, \qquad
    L_{33}=(q_0\,q_0^*-q_1\,q_1^* )-(q_2\,q_2^*-q_3\,q_3^*) \, ;
        \nonumber\label{1.12b}\end{eqnarray}
\textbf{block} $(N)$\begin{eqnarray}
    L_{02}=- (q_0\,q_2^*+q_2\,q_0^* )
        -i\,(q_1\,q_3^*-q_3\,q_1^*)\,,\qquad L_{03}=-
        (q_0\,q_3^*+q_3\,q_0^* ) +i\,(q_1\,q_2^*-q_2\,q_1^*) \,,\nonumber\\
    L_{12}=i\,(q_0\,q_3^* -q_3\,q_0^*)+(q_1\,q_2^*+q_2\,q_1^* )  \,,\qquad
    L_{13}=-i\,(q_0\,q_2^*  -q_2\,q_0^*)+(q_1\,q_3^*+q_3\,q_1^* ) \,;
        \nonumber\label{1.12c}\end{eqnarray}
\textbf{block} $(L)$\begin{eqnarray}
    L_{20}=-(q_0\,q_2^*+q_2\,q_0^*)+i\,(q_1\,q_3^*-q_3\,q_1^*)\,,\qquad
    L_{21}=-i\,(q_0\,q_3^*-q_3\,q_0^*)+(q_1\,q_2^*+q_2\,q_1^*) \,,\nonumber\\
    L_{30}=-(q_0\,q_3^*+q_3\,q_0^*)-i\,(q_1\,q_2^*-q_2\,q_1^*)\,,\qquad
    L_{31}=i\,(q_0\,q_2^*-q_2\,q_0^*)+(q_1\,q_3^*+q_3\,q_1^*) \, .
        \nonumber\label{1.12d}\end{eqnarray}
A definite ordering structure will be seen in the matrix $L$, if one decomposes $L$
    into symmetric $S$ and skew-symmetric $A$ parts:\footnote{Cf. \cite{1}, $L=S+A$, with $S=S^t,A=-A^t$.}
    \begin{eqnarray}S_{00}=L_{00} =q_0\,q_0^*+q_1\,q_1^*+q_2\,q_2^*+q_3\,q_3^*\;,\qquad
        S_{11}=L_{11}=q_0\,q_0^*+q_1\,q_1^* -q_2\,q_2^* -q_3\,q_3^*\;,\nonumber\\
    S_{22}=  q_0\,q_0^*+q_2\,q_2^*-q_1\,q_1^*   -q_3\,q_3^*\;,\qquad
    S_{33}=q_0\,q_0^*+q_3\,q_3^*-q_1\,q_1^* -q_2\,q_2^*\;,\nonumber\end{eqnarray}
    \begin{eqnarray}S_{01}=- (q_0 q_1^*+q_0^* q_1)\;,\;\;
    S_{02} =- (q_0 q_2^*+q_0^* q_2)\;,\;\;
    S_{03} =- (q_0 q_3^*+q_0^* q_3)\;,\nonumber\\
    S_{12}= q_1q_2^*+q_1^*q_2 \;,\qquad
    S_{13}=q_1q_3^*+q_1^*q_3 \;,\qquad
    S_{23}= q_2q_3^*+q_2^*q_3 \;, \nonumber\end{eqnarray}
    \begin{eqnarray}A_{01}= i ( q_2 q_3^*-q_2^* q_3)\;,
    \;\;A_{02} =i ( q_3 q_1^*-q_3^* q_1)\;,\;\;
    A_{03}= i ( q_1 q_2^*-q_1^* q_2)\;,\nonumber\\
    A_{12}=-i (q_3 q_0^*- q_3^* q_0 ) \;,\;\;
    A_{13}=+i (q_2 q_0^*- q_2^* q_0 ) \;,\;\;
    A_{23}= -i (q_1 q_0^*- q_1^* q_0 ) \; .\nonumber\end{eqnarray}
We easily get
    \begin{eqnarray}Q_0^2=k_0k_0^*={1 \over 4}
        \mbox{Sp}\; L\; ={S_{00}+S_{11}+S_{22}+S_{33} \over 4}\;,\nonumber\\
    Q_1^2=q_1q_1^*={S_{00}+S_{11} \over 2}- {1 \over 4} \mbox{Sp}\;
        L={S_{00}+S_{11}-S_{22}-S_{33} \over 4}\;,\nonumber\\
    Q_2^2=  q_2q_2^*={S_{00}+S_{22} \over 2}- {1 \over 4}
        \mbox{Sp}\; L={S_{00}+S_{22}-S_{11}-S_{33} \over 4} \;,\nonumber\\
    Q_3^2= q_3q_3^*={S_{00}+S_{33} \over 2}- {1 \over 4}
        \mbox{Sp}\; L={S_{00}+S_{33}-S_{11}-S_{22} \over 4} \; .\label{1.14}\end{eqnarray}
Then, by means of the identities
    \begin{eqnarray}S_{01}+ i A_{23}=- 2 q_0 q_1^* \;, \qquad S_{23}+i
        A_{01}=2q_2^* q_3 \;,\nonumber\\
    S_{02}+ i A_{31}=- 2 q_0 q_2^* \;,\qquad S_{31}+iA_{02}=2q_3^* q_1 \;,\nonumber\\
    S_{03}+ i A_{12}=- 2 q_0 q_3^* \;, \qquad  S_{12}+iA_{03}=2q_1^*q_2 \; .\label{1.15a}\end{eqnarray}
or, using the complex polar representation,\footnote{Here, we denote
    $Q_k=\,\mbox{abs}\,(q_k)$, and $\alpha_k=arg(q_k)$.}
    \begin{eqnarray}S_{01}+ i A_{23}=- 2 Q_0 Q_1 e^{i(\alpha_0-\alpha_1 )} \;,\qquad
    S_{23}+i A_{01}=2 Q_2 Q_3e^{i(-\alpha_2+\alpha_3 ) } \;,\nonumber\\
    S_{02}+ i A_{31}=- 2 Q_0 Q_2 e^{i(\alpha_0-\alpha_2 )} \;,\qquad
    S_{31}+i A_{02}=2 Q_3 Q_1e^{i(-\alpha_3+\alpha_1 ) } \;,\nonumber\\
    S_{03}+ i A_{12}=- 2 Q_0 Q_3 e^{i(\alpha_0-\alpha_3 )} \; ,\qquad
    S_{12}+i A_{03}=2 Q_1 Q_2e^{i(-\alpha_1+\alpha_2 ) } \; .\label{1.15b}\end{eqnarray}
Using three equations from the above ones, we express the phases $\alpha_1,\alpha_2,\alpha_3$
    through $\alpha_0$:
    \begin{eqnarray}e^{i\alpha_1} =- {2 Q_0 Q_1  \over S_{01}+ i A_{23}}e^{i\alpha_0}
    \qquad  \Longrightarrow \qquad q_1 =- {2  Q_1^2 \over S_{01}+ i A_{23}} \; q_0 \;,\nonumber\\
    e^{i\alpha_2} =- {2 Q_0 Q_2 \over S_{02}+ i A_{31}}
    e^{i\alpha_0} \qquad  \Longrightarrow \qquad
    q_2 =- {2  Q_2^2 \over S_{02}+ i A_{31}} \; q_0 \;,\nonumber\\
    e^{i\alpha_3}=- {2 Q_0 Q_3  \over S_{03}+ i A_{12} }
    e^{i\alpha_0 } \qquad  \Longrightarrow \qquad
    q_3 =- {2  Q_3^2 \over S_{03}+ i A_{12}} \; q_0\; ,\label{1.16}\end{eqnarray}
Now, it remains to consider the basic restriction on parameters $q_0^2-{\bf q}^2 =+1 \; ;$ so we get
    \begin{eqnarray}q_0^2 \; \left [ 1   -   {4  Q_1^{4}  \over (S_{01}+ i A_{23})^2 }-
        {4 Q_2^{4} \over (S_{02}+ i A_{31})^2 } -{4  Q_3^{4}  \over (S_{03}+ i A_{12})^2  }
        \right ] =+1 \; .\label{1.17b}\end{eqnarray}
Thus, we find  $q_0$:
    \begin{eqnarray}q_0=  \pm \left [ 1   -   {4  Q_1^{4}  \over (S_{01}+
        i A_{23})^2 }-{4 Q_2^{4} \over (S_{02}+ i A_{31})^2 } -{4  Q_3^{4}
        \over (S_{03}+ i A_{12})^2  } \right ]^{-1/2 } \; .\label{1.18a}\end{eqnarray}
and then arrive at expression of $q_{j}$ in terms of $q_0, S_{0j},A_{ij}$ and $Q_j$ ($j=\overline{1,3}$):
    \begin{eqnarray}q_1 =- {2  Q_1^2 \over S_{01}+ i A_{23}} \; q_0 \;,\qquad
    q_2 =- {2  Q_2^2 \over S_{02}+ i A_{31}} \; q_0 \;,\qquad
    q_3 =- {2  Q_3^2 \over S_{03}+ i A_{12}} \; q_0\;  .\label{1.18b}\end{eqnarray}
%
\section{On identifying Lorentz--Mueller matrices within the
total linear group $GL(4,R)$}
\hspace{5mm}The Mueller matrices of Lorentzian type $M=L$ form a real subgroup of the
    total linear group $GL(4,R)$
    \begin{eqnarray}G = \left(\begin{array}{rr}k_0+\; {\bf k} \; \vec{\sigma}  \;\;&
        n_0+\; {\bf n}  \; \vec{\sigma}\\[3mm]
    l_0+\; {\bf l} \; \vec{\sigma} \;\;&m_0+\; {\bf m}\; \vec{\sigma}\end{array} \right),
        \label{2.1a}\end{eqnarray}
where $k_0\equiv k_0I_2$ and $\sigma_0=I_2$, ${\bf k}=(k_1,k_2,k_3)$ and
    ${\vec\sigma}=(\sigma_1,\sigma_2,\sigma_3)$. To any $G$ of the form (\ref{2.1a}),
    there corresponds a natural invariant, its determinant
    (whose explicit form in such parametrization was given in \cite{10}):
\begin{eqnarray}\mbox{det}\; G=(kk) \; (mm)+(nn)\;(ll)-2\; (kn)\;(ml)\;  -\nonumber\\
    2 \;( \;-k_0\;{\bf n}+ n_0 \; {\bf k}+ i \;{\bf k}  \times  {\bf n} \;) \;
    (\; -m_0 \;{\bf l}+l_0\;{\bf m}+i\; {\bf m}  \times  {\bf l} \; )   \;,\label{2.1b}\end{eqnarray}
where the notation $(kk)$ means $(kk)= k_0^2-{\bf k}^2$, etc.

In \cite{6,7,8}, a special classification for degenerate matrices with vanishing determinant was developed;
    only part of them may be of Mueller type. Such a class of degenerate matrices is not considered  here.
Considering the above Lorentz--Mueller matrix as consisting of four blocks
    \begin{eqnarray}L (q,q^*)=\left(\begin{array}{cccc}
    k_0+k_3&  k_1-i k_2&n_0+n_3& n_1 -i n_2\\
    k_1+i  k_2&k_0-k_3&  n_1+i n_2&n_0-n_3\\
    l_0+l_3&  l_1 -i l_2&  m_0+m_3& m_1 -i m_2\\
    l_1+i  l_2&l_0-l_3&  m_1+i m_2&m_0-m_3
    \end{array} \right)=\left(\begin{array}{cc}K&N\\L&M\end{array} \right),\label{2.2}\end{eqnarray}
we can find the associated 16 coefficients $k,m,n,l$ by fixing an arbitrary matrix
    $L(q,q^*)$. We shall use the explicit form for elements of the matrix $L$.
    It is convenient to introduce the change of variables
    $$k_2 \Longrightarrow i k_2, \qquad m_2 \Longrightarrow i m_2, \qquad
        n_2 \Longrightarrow i n_2, \qquad  l_2 \Longrightarrow i l_2\; .$$
After some simple calculation, we get
    $$\begin{array}{ll}k_0= q_0\,q_0^*+q_1\,q_1^*\,,&
        k_1=-(q_0\,q_1^*+q_1\,q_0^*)\,,\smallskip\\
    k_2=i\,(q_2\,q_3^*-q_3\,q_2^*)\,,&
        k_3=q_2\,q_2^*+q_3\,q_3^*\,,\smallskip\\
    m_0=q_0\,q_0^*-q_1\,q_1^*\,,&
        m_1=q_2\,q_3^*+q_3\,q_2^*\,,\smallskip\\
    m_2=i\,(q_0\,q_1^* -q_1\,q_0^*)\,,&
        m_3=-q_2\,q_2^* -q_3\,q_3^*\,,\end{array}$$
    \begin{equation}\label{2.3}\begin{array}{l}
    2\,l_0=-(q_0\,q_2^*+q_2\,q_0^*)+i\,(q_0\,q_2^*-q_2\,q_0^*)
       +i\,(q_1\,q_3^* -q_3\,q_1^*)+(q_1\,q_3^*+q_3\,q_1^*)\,,\\
    2\,l_3=-(q_0\,q_2^*+q_2\,q_0^*)-i\,(q_0\,q_2^*-q_2\,q_0^*)
       + i\,(q_1\,q_3^* -q_3\,q_1^*) -(q_1\,q_3^*+q_3\,q_1^*)\,,\\
    2\,l_1=-i\,(q_0\,q_3^*-q_3\,q_0^*)-(q_0\,q_3^*+q_3\,q_0^*)
       +(q_1\,q_2^*+q_2\,q_1^*) -i\,(q_1\,q_2^*-q_2\,q_1^*) \,,\\
    2\,l_2=-i\,(q_0\,q_3^*-q_3\,q_0^*)+ (q_0\,q_3^*+q_3\,q_0^*)
       +(q_1\,q_2^*+q_2\,q_1^*)+i\,(q_1\,q_2^*-q_2\,q_1^*)\,,\bigskip\\
    2n_0=-(q_0\,q_2^*+q_2\,q_0^*)-i\,(q_0\,q_2^*-q_2\,q_0^*)
        -i\,(q_1\,q_3^* -q_3\,q_1^*)+(q_1\,q_3^*+q_3\,q_1^*)\,,\\
    2n_3=-(q_0\,q_2^*+q_2\,q_0^*)+i\,(q_0\,q_2^*-q_2\,q_0^*)
        -i\,(q_1\,q_3^* -q_3\,q_1^*) -(q_1\,q_3^*+q_3\,q_1^*) \,,\\
    2n_1=- (q_0\,q_3^*+q_3\,q_0^* )+i\,(q_0\,q_3^* -q_3\,q_0^*)
       +i\,(q_1\,q_2^*- q_2\,q_1^*)+( q_1\,q_2^*+q_2\,q_1^* )\;,\\
    2n_2=-(q_0\,q_3^*+q_3\,q_0^* ) -i\,(q_0\,q_3^* -q_3\,q_0^*)
       +i\,(q_1\,q_2^*- q_2\,q_1^*) -( q_1\,q_2^*+q_2\,q_1^* )\; .
    \end{array}\end{equation}
\section{ On the expansion of the Lorentz matrices in Dirac
basis}
\hspace{5mm}We can alternatively rephrase the previous problem: one can decompose the Lorentz matrix in
    terms of the 16 Dirac matrices
    \begin{eqnarray}L(q,q^*)= Z+\gamma^5 \; \tilde{Z }+\gamma^l \; Z_{l}+\gamma^l \gamma^5 \;
        \tilde {Z }_{l} +\sigma^{mn}\;  Z _{mn}   \; ;\label{3.1a}\end{eqnarray}
where the 16 coefficients are given by the
    formulas\footnote{We denote by $Sp$ the trace of the corresponding matrix.}
    \begin{equation}\label{3.1b}\begin{array}{c}Z={1 \over 4} \; \mbox{Sp}\; L(q,q^*) \;, \qquad
        \tilde{Z}={1 \over 4} \; \mbox{Sp}\; \gamma^5 L(q,q^*) \;,\medskip\\
    Z_{k}={1 \over 4}\; \mbox{Sp}\; \gamma_{k} L(q,q^*) \;, \qquad
        \tilde{Z}_{k}={1 \over 4}\; \mbox{Sp}\; \gamma^5 \gamma_{k} L(q,q^*) \;,\qquad
    Z_{kl}=-{1 \over 2} \; \mbox{Sp}\; \sigma_{kl} L(q,q^*) \;.\end{array}\end{equation}
After simple calculations we arrive at the formulas
\begin{eqnarray}
Z= q_0\,q_0^* \;, \qquad \tilde{Z}=q_1\,q_1^*\;, \qquad
Z_{03}=q_3\,q_3^*\;, \qquad
-i Z_{12}=q_2\,q_2^* \;,
\nonumber
\\
Z_{01} =-{1\over 2}\, [ (q_0\,q_1^*+q_1\,q_0^* )+(q_2\,q_3^*+q_3\,q_2^*) ]\,,\qquad
Z_{23} ={i\over 2}\,[-( q_0\,q_1^*+ q_1\,q_0^*)+(q_2\,q_3^*+q_3\,q_2^*) ]\;,
\nonumber
\\
Z_{02} ={1\over 2}\,[ (q_0\,q_1^*
-q_1\,q_0^* )-(q_2\,q_3^*-q_3\,q_2^*)] \,,\qquad
Z_{31}={i\over 2}\, [-(q_0\,q_1^* -q_1\,q_0^*)-(q_2\,q_3^* -q_3\,q_2^*)]
\;,
\nonumber
\\
Z_0={1 \over 2}\, [ -(q_0\,q_2^*+q_2\,q_0^*)+(q_1\,q_3^*+q_3\,q_1^*)]\,,\qquad
\tilde{Z}_0= {i \over 2}\,
[+(q_0\,q_2^* -q_2\,q_0^*)+(q_1\,q_3^* -q_3\,q_1^*) ] \,,
\nonumber
\\
Z_3=-{i \over 2}\, [   (q_0\,q_2^* -q_2\,q_0^* )-(q_1\,q_3^* -q_3\,q_1^*)] \,,\qquad
\tilde{Z}_3=
-{1 \over 2}\,[  (q_0\,q_2^*+q_2\,q_0^* )+(q_1\,q_3^*+q_3\,q_1^*)] \,,
\nonumber
\\
Z_1 =-{i\over 2}\,(q_0\,q_3^* -q_3\,q_0^*+q_1\,q_2^* -q_2\,q_1^*)\,,\qquad
\tilde{Z}_1= -{1\over 2}\, [(q_0\,q_3^*+q_3\,q_0^* )-(q_1\,q_2^*+
q_2\,q_1^*) ]\,,
\nonumber
\\
Z_2=-{i\over 2}\,(-q_0\,q_3^* -q_3\,q_0^*-q_1\,q_2^* -q_2\,q_1^*)\,,\qquad
\tilde{Z}_2= -{1\over 2}\,[-(q_0\,q_3^* -q_3\,q_0^*)+(q_1\,q_2^* -q_2\,q_1^*)]\,.
\nonumber
\\
\label{3.3}
\end{eqnarray}
On the ground of the formulas (\ref{3.3}), one can develop a method to find
    the parameters $q_{a}$ by means of the coefficients in (\ref{3.1a}).
    Indeed,  starting  with   (\ref{3.3}), we note
\begin{eqnarray}
Z= q_0\,q_0^* \;, \qquad \tilde{Z}=q_1\,q_1^*\;, \qquad
Z_{03}=q_3\,q_3^*\;, \qquad
-i Z_{12}=q_2\,q_2^* \; ;
\label{3.4a}
\end{eqnarray}
then
$$
Z_{01}+ i Z_{23}=-q_2 q_3^*-q_2^* q_3 \;,
\qquad
Z_{01}-i Z_{23}=- q_0 q_1^*-q_0^* q_1 \;,
$$
$$
Z_{02}+ i Z_{31}  =q_0 q_1^*-q_0^* q_1 \;, \qquad
Z_{02}-i Z_{31}=
-q_2q_3^*+q_2^*q_3\;,
$$
that is
\begin{eqnarray}
(Z_{01}-i Z_{23})+(Z_{02}+ i Z_{31})=-2 \; q_0^* q_1 \;,
\nonumber
\\
(Z_{01}-i Z_{23})-(Z_{02}+ i Z_{31})=-2 \; q_0 q_1^* \;,
\nonumber
\\
(Z_{01}+ i Z_{23})+(Z_{02}-i Z_{31})=-2 \; q_2 q_3^* \;,
\nonumber
\\
Z_{01}+ i Z_{23})- (Z_{02}-i Z_{31})=-2 \; q_2^* q_3 \; ;
\label{3.4b}
\end{eqnarray}
then
\begin{eqnarray}
Z_0-\tilde{Z}_3=(q_1q_3^*+q_1^* q_3)\;,\qquad
Z_0+\tilde{Z}_3=- (q_0 q_2^*+ q_0^*q_2)\;,
\nonumber
\\
\tilde{Z}_0-Z_3=i (q_0 q_2^*-q_0^* q_2)\;,  \qquad
\tilde{Z}_0+Z_3=i (q_1 q_3^*-q_1^* q_3)\;,
\nonumber
\end{eqnarray}
that is
\begin{eqnarray}
(Z_0-\tilde{Z}_3 )+i (\tilde{Z}_0+Z_3)= 2 \; q_1^* q_3 \;,
\nonumber
\\
(Z_0-\tilde{Z}_3 )-i (\tilde{Z}_0+Z_3)= 2 \; q_1 q_3^* \; ;
\nonumber
\\
(Z_0+\tilde{Z}_3)+i (\tilde{Z}_0-Z_3)=-2 \; q_0q_2^* \;,
\nonumber
\\
(Z_0+\tilde{Z}_3)-i (\tilde{Z}_0-Z_3)= -2 \; q_0^*q_2 \; ;
\label{3.4c}
\end{eqnarray}
and then
\begin{eqnarray}
Z_1+i \tilde{Z}_2=-i (q_1 q_2^*-q_1^*q_2)\;, \qquad
Z_1-i \tilde{Z}_2=-i (q_0 q_3^*- q_0^*q_3) \;,
\nonumber
\\
\tilde{Z}_1+i Z_2==q_0q_3^*-q_0^* q_3 \;, \qquad
\tilde{Z}_1-i Z_2= q_1q_2^*+q_1^*q_2 \; .
\nonumber
\end{eqnarray}
that is
\begin{eqnarray}
(Z_1+i \tilde{Z}_2)+i (\tilde{Z}_1-i Z_2)=+ 2i \; q_1^*q_2 \;,
\nonumber
\\
(Z_1+i \tilde{Z}_2)-i (\tilde{Z}_1-i Z_2)=-2i\;  q_1 q_2^* \;,
\nonumber
\\
(Z_1-i \tilde{Z}_2)+i (\tilde{Z}_1+i Z_2)= -2i \; q_0q_3^*   \;,
\nonumber
\\
(Z_1-i \tilde{Z}_2) -i (\tilde{Z}_1+i Z_2)= +2i \; q_0^*q_3   \; .
\label{3.4d}
\end{eqnarray}
The previously produced formulas allow us to calculate the parameters for Lorentz matrices.
    Indeed, let us take relations
\begin{eqnarray}Z= q_0\,q_0^* \qquad q_0=\sqrt{Z}e^{i\alpha} \;,\nonumber\\
  q_1 =- {1 \over  2 \; q_0^*}\;  [ (Z_{01}-i Z_{23})+(Z_{02}+ i Z_{31}) ]={1 \over q_0^*} M_1 \;,\nonumber\\
 q_2=-{1 \over 2 \;  q_0^*} \; [ (Z_0+\tilde{Z}_3)-i (\tilde{Z}_0-Z_3)  ]={1 \over q_0^*} M_2\;,\nonumber\\
q_3 =-i {1 \over 2 \; q_0^* }\;  [ (Z_1-i \tilde{Z}_2) -i (\tilde{Z}_1+i Z_2) ]={1 \over q_0^*} M_3     \; .
\label{3.5a}
\end{eqnarray}
With the us of additional quadratic restriction $q_0^2-{\bf q}^2=1$, we derive the formulas
\begin{eqnarray}
e^{i\alpha}=\pm \sqrt{ {Z \over  Z^2- {\bf M}^2 }}, \qquad q_0=\sqrt{Z}  e^{i\alpha}\;, \qquad
q_{j}={e^{i \alpha}  \over \sqrt{Z}}  M_{j}\; .
\label{3.5b}
\end{eqnarray}
The expansion of the Lorentz matrices in the Dirac basis can be written
as
\begin{eqnarray}L=\left(\begin{array}{cc}Z+\tilde{Z}+\Sigma^{mn} Z_{mn}&\bar{\sigma}^{n} (Z_{n}  -\tilde{Z}_{n} \\[3mm]
  \sigma^{n}( Z_{n} +\tilde{Z}_{n})&Z-\tilde{Z}+\bar{\Sigma}^{mn} Z_{mn}
\end{array} \right)\label{3.7}\end{eqnarray}
where
\begin{eqnarray}
\sigma^{a}=(I,+\sigma^{j}) \;, \qquad \sigma_{a}=(I,-\sigma^{j}) \;,\nonumber\\
\gamma^{a}=\left(\begin{array}{cc}
0&\bar{\sigma}^{a}\\\sigma^{a}&0\end{array} \right), \qquad
\left(\begin{array}{cc}
+I&0\\
0&-I
\end{array} \right)
= \gamma^5, \qquad
 \sigma^{ab}={1 \over 4} (\gamma^{a} \gamma^{b}-\gamma^{b}
\gamma^{a} )=\left(\begin{array}{cc}
\Sigma^{ab}&0\\
0&\bar{\Sigma}^{ab}
\end{array}
\right), \nonumber
\\
\Sigma^{01}={1 \over 4}(  \bar{\sigma}^{0} \sigma^{1}-\bar{\sigma}^{1} \sigma^{0} )={1 \over 2} \sigma^{1} \; , \qquad
\Sigma^{02}={1 \over 2} \sigma^2 \;, \qquad \Sigma^{03}={1
\over 2} \sigma^{3} \;,
\nonumber
\\
\Sigma^{12}={1 \over 4}(  \bar{\sigma}^{1} \sigma^2-\bar{\sigma}^2 \sigma^{1} )=-{i \over 2} \sigma^{3} \; , \qquad
\Sigma^{23}=-{i \over 2} \sigma^{1} \; ,\qquad \Sigma^{31}=-{i \over 2} \sigma^2 \;  ;
\nonumber
\\
\bar{\Sigma}^{01}={1 \over 4}(  \sigma^{0} \bar{\sigma}^{1}-\sigma^{1} \bar{\sigma}^{0} )=-{1 \over 2} \sigma^{1} \; , \qquad
\bar{\Sigma}^{02}=-{1 \over 2} \sigma^2 \;, \qquad
\bar{\Sigma}^{03}=-{1 \over 2} \sigma^{3} \;,
\nonumber
\\
\bar{\Sigma}^{12}={1 \over 4}(  \sigma^{1} \bar{\sigma}^2-\sigma^2 \bar{\sigma}^{1} )=-{i \over 2} \sigma^{3} \; , \qquad
\bar{\Sigma}^{23}=-{i \over 2} \sigma^{1} \; ,\qquad
\bar{\Sigma}^{31}=-{i \over 2} \sigma^2 \;  .
\label{3.8}
\end{eqnarray}
Relation  (\ref{3.7})  reduces to the form
\begin{eqnarray}
L=\left(\begin{array}{cc}
Z+\tilde{Z}+\sigma^{1} Z^{-} _1+  \sigma^2 Z^{-} _2+ \sigma^{3} Z^{-} _3&Z_0  -\tilde{Z}_0-\sigma^{j} (Z_{j}  -\tilde{Z}_{j} ) \\[3mm]
 Z_0 +\tilde{Z}_0+\sigma^{j} (Z_{j} +\tilde{Z}_{j} )&Z-\tilde{Z}- \sigma^{1} Z^{+} _1 -\sigma^2 Z^{+} _2-
\sigma^{3} Z^{+} _3
\end{array} \right),
\label{3.9a}
\end{eqnarray}
where
\begin{eqnarray}
Z^{-} _1=+Z_{01}-i Z_{23} , \qquad Z^{-}_2=+Z_{02}-i Z_{31}, \qquad Z^{-}_3=+Z_{03}-i Z_{12} \;,
\nonumber
\\
Z^{+} _1=Z_{01}+i Z_{23} , \qquad Z^{+}_2=Z_{02}+i Z_{31}, \qquad Z^{+}_3= Z_{03}+i Z_{12} \; .
\label{3.9b}
\end{eqnarray}
The matrix (\ref{3.9a}) can be considered as consisting of four blocks
\begin{eqnarray}
L=\left( \begin{array}{cc}
k_0+k_{j} \sigma^{j}&n_0+n_{j} \sigma^{j}\\
l_0+l_{j} \sigma^{j}&m_0+m_{j} \sigma^{j}
\end{array} \right);
\label{3.10a}
\end{eqnarray}
and explicit expressions for the parameters are
\begin{eqnarray}
k_0=Z+\tilde{Z}  \;, \qquad m_0=Z -\tilde{Z} \;,
\qquad
k_{j}=+Z^{-} _{j}\;, \qquad   m_{j}=-Z^{+} _{j} \;,
\nonumber
\\
n_0=Z_0  -\tilde{Z}_0 \;, \qquad l_0=Z_0 +\tilde{Z}_0 \;,
\qquad
n_{j}=-Z_{j} +\tilde{Z}_{j} \;, \qquad
l_{j}=  Z_{j} +\tilde{Z}_{j}  \; ;
\label{3.10b}
\end{eqnarray}
and it can be readily verified that these coincide with (\ref{2.2}).
\section{On parameters of Lorentz--Mueller matrices and
transitivity relations}

\hspace{5mm}
Let us start with  the factorized form of Lorentz transformations
\begin{eqnarray}A A^*=L  \qquad \Longrightarrow \qquad A=L  (A^*)^{-1} \;,
\nonumber\\
 A^*  A=L  \qquad \Longrightarrow \qquad A^*=L  A^{-1} \; .
\label{5.1}
\end{eqnarray}
whence we get
\begin{eqnarray}
L \; [ (A^*)^{-1}+  A^{-1} ]= A+A^*  \;, \qquad
 L  \; [ (A^*)^{-1}-  A^{-1} ]=A-A^* \;,
\label{5.2}
\end{eqnarray}
where
\begin{eqnarray}
A =\left(\begin{array}{rrrr}
 q_0&-q_1&-q_2&-q_3\\
-q_1& q_0&-iq_3&iq_2\\
-q_2&iq_3&q_0&-iq_1\\
-q_3&-iq_2&iq_1&q_0
\end{array}  \right), \qquad
A^*=\left(\begin{array}{rrrr}
 q_0^*&-q_1^*&-q_2^*&-q_3^*\\
-q_1^*&q_0^*&iq_3^*&-iq_2^*\\
-q_2^*&-iq_3^*&q_0^*&iq_1^*\\
-q_3^*&iq_2^*&-iq_1^*&q_0^*
\end{array}  \right),
\nonumber
\\
A^{-1}  =\left(\begin{array}{rrrr}
 q_0&q_1&q_2&q_3\\
q_1& q_0&iq_3&-iq_2\\
q_2&-iq_3&q_0&iq_1\\
q_3&iq_2&-iq_1&q_0
\end{array}  \right), \qquad
(A^* )^{-1}=\left(\begin{array}{rrrr}
 q_0^*&q_1^*&q_2^*&q_3^*\\
q_1^*&q_0^*&-iq_3^*&iq_2^*\\
q_2^*&iq_3^*&q_0^*&-iq_1^*\\
q_3^*&-iq_2^*&iq_1^*&q_0^*
\end{array}  \right).
\nonumber
\end{eqnarray}
For the special case of Euclidean rotation
$
q_0=n_0  \;, \;  q_{j}=  -i  n_{j}$,
the relations (\ref{5.2}) give
$$
\left(\begin{array}{cccc}
1&0&0&0\\
0&L_{11}&L_{12}&L_{13}\\
0&L_{21}&L_{22}&L_{23}\\
0&L_{31}&L_{32}&L_{33}
\end{array} \right) \left(\begin{array}{cccc}
n_0&0&0&0\\
0&n_0&n_3&-n_2\\
0&-n_3&n_0&n_1\\
0&n_2&-n_1&n_0
\end{array} \right)=\left(\begin{array}{cccc}
n_0&0&0&0\\
0&n_0&-n_3&n_2\\
0&n_3&n_0&-n_1\\
0&-n_2&n_1&n_0
\end{array} \right)\;,
$$
$$
\left(\begin{array}{cccc}
1&0&0&0\\
0&L_{11}&L_{12}&L_{13}\\
0&L_{21}&L_{22}&L_{23}\\
0&L_{31}&L_{32}&L_{33}
\end{array} \right)
\left(\begin{array}{cccc}
0&in_1&i n_2&i n_3\\
i n_1&0&0&0\\
i n_2&0&0&0\\
i n_3&0&0&0
\end{array} \right)=\left(\begin{array}{cccc}
0&in_1&i n_2&i n_3\\
i n_1&0&0&0\\
in_2&0&0&0\\
i n_3&0&0&0
\end{array} \right) .
$$
In fact, we have four transitivity relations
\begin{eqnarray}
\left(\begin{array}{ccc}
 L_{11}&L_{12}&L_{13}\\
 L_{21}&L_{22}&L_{23}\\
 L_{31}&L_{32}&L_{33}
\end{array} \right) \;
\left(\begin{array}{r}
  n_0\\
  -n_3\\
 n_2
\end{array}  \right)=
\left(\begin{array}{r}
   n_0\\
  n_3\\
  -n_2
\end{array}  \right),
\nonumber
\\
\left(\begin{array}{ccc}
L_{11}&L_{12}&L_{13}\\
 L_{21}&L_{22}&L_{23}\\
 L_{31}&L_{32}&L_{33}
\end{array} \right) \;
\left(\begin{array}{r}
n_3\\
  n_0\\
- n_1
\end{array}  \right)=\left(\begin{array}{r}
 -n_3\\
  n_0\\
 n_1
\end{array}  \right),
\nonumber
\\
\left(\begin{array}{ccc}
 L_{11}&L_{12}&L_{13}\\
 L_{21}&L_{22}&L_{23}\\
 L_{31}&L_{32}&L_{33}
\end{array} \right) \;
\left(\begin{array}{r}
   -n_2\\
n_1\\
   n_0
\end{array}  \right)= \left(\begin{array}{r}
   n_2\\
-n_1\\
   n_0
\end{array}  \right),
\nonumber
\\
\left(\begin{array}{ccc}
 L_{11}&L_{12}&L_{13}\\
 L_{21}&L_{22}&L_{23}\\
 L_{31}&L_{32}&L_{33}
\end{array} \right)
\left(\begin{array}{c}
 n_1\\
n_2\\
 n_3
\end{array} \right)= \left(\begin{array}{c}
 n_1\\
 n_2\\
 n_3
\end{array} \right),
\label{5.4b}
\end{eqnarray}
where
\begin{eqnarray}
O       =\left(\begin{array}{lll}
 1 -2 (n_2^2+n_3^2)&-2n_0n_3+2n_1n_2&+2n_0n_2+2n_1n_3\\
+2n_0n_3+2n_1n_2&1 -2 (n_3^2+n_1^2)&-2n_0n_1+2n_2n_3\\
 -2n_0n_2+2n_1n_3&+2n_0n_1+2n_2n_3&1 -2 (n_1^2+n_2^2)
 \end{array} \right),\;\;
 n_0^2+ n_1^2+n_2^2+n_3^2 =1  \; .
\label{5.4c}
\end{eqnarray}
It can be readily verified that these four relations are valid indeed. In the context
    of the problems of polarization optics,  the most interesting is the last one from (\ref{5.4b})
\begin{eqnarray}
O \left(\begin{array}{c}
 n_1\\
n_2\\
 n_3
\end{array} \right)= \left(\begin{array}{c}
 n_1\\
 n_2\\
 n_3
\end{array} \right),
\label{5.4d}
\end{eqnarray}
However, all the four relations allow physical interpretation in the frames of polarization optics:
    they provide us with simple transitivity relations that describe the action of an optical element on
    a specially chosen probe of light beams.

In a similar manner, one may consider the case of pseudo-Euclidean rotations
    $q_0= m_0 \;, \;   q_{j}=  m _{j}$, whence (\ref{5.2}) take the form
    \begin{eqnarray}\left(\begin{array}{rrrr}L_{00}&L_{01}&L_{02}&L_{03}\\
        L_{10}&L_{11}&L_{12}&L_{13}\\L_{20}&L_{21}&L_{22}&L_{23}\\
        L_{30}&L_{31}&L_{32}&L_{33}\end{array} \right)
    \left(\begin{array}{rrrr} m_0&m_1&m_2&m_3\\m_1& m_0&0&0\\m_2&0&m_0&0\\
        m_3&0&0&m_0\end{array}  \right)=
    \left(\begin{array}{rrrr} m_0&-m_1&-m_2&-m_3\\-m_1& m_0&0&0\\-m_2&0&m_0&0\\
        -m_3&0&0&m_0\end{array}  \right),\nonumber\\
    \left(\begin{array}{rrrr}L_{00}&L_{01}&L_{02}&L_{03}\\L_{10}&L_{11}&L_{12}&L_{13}\\
        L_{20}&L_{21}&L_{22}&L_{23}\\L_{30}&L_{31}&L_{32}&L_{33}\end{array} \right)
    \left(\begin{array}{rrrr} 0&0&0&0\\0& 0&-im_3&im_2\\0&im_3&0&-im_1\\0&-im_2&im_1&0\end{array}\right)=
    \left(\begin{array}{rrrr} 0&0&0&0\\0& 0&-im_3&im_2\\0&im_3&0&-im_1\\0&-im_2&im_1&0\end{array}\right).
    \label{5.5c}\end{eqnarray}
Using the explicit formulas
    $$m_0=\mbox{ch}\; \chi\;,\qquad  m_{j}=\mbox{sh}\; \chi e_{j},$$
where ${\bf e}^2=1$ which infer $m_0^2-{\bf m}^2=1$, we get
    \begin{equation}\label{5.6a}
    L =\left(\begin{array}{cccc}m_0^2+{\bf m}^2&-2m_0m_1&-2m_0m_2&-2m_0m_3\\
        -2m_0m_1&1+2 m_1^2&2m_1m_2&2m_1m_3\\-2m_0m_2&2m_1m_2&1 +2m_2^2&2m_2m_3\\
        -2m_0m_3&2m_3m_1&2m_2m_3&1 +2m_3^2\end{array} \right),\end{equation}
and one can easily verify that such an equation holds indeed
    \begin{eqnarray}
    L \left(\begin{array}{rrrr}
        m_0&m_1&m_2&m_3\\m_1& m_0&0&0\\m_2&0&m_0&0\\m_3&0&0&m_0\end{array}  \right)=
    \left(\begin{array}{rrrr}
        m_0&-m_1&-m_2&-m_3\\-m_1& m_0&0&0\\-m_2&0&m_0&0\\-m_3&0&0&m_0\end{array}\right),
        \nonumber\\
    L \left(\begin{array}{rrrr}0&0&0&0\\0& 0&-im_3&im_2\\0&im_3&0&-im_1\\
        0&-im_2&im_1&0\end{array}  \right)=
    \left(\begin{array}{rrrr} 0&0&0&0\\0& 0&-im_3&im_2\\0&im_3&0&-im_1\\
        0&-im_2&im_1&0\end{array}  \right).\label{5.6c}\end{eqnarray}
Thus, there arise 7 non-trivial  transitivity relations.
%
Among the 7 vectors from (\ref{5.6c}), only one (the first) is
time-similar
    \begin{eqnarray}m_0^2-m_1^2-m_2^2-m_3^2=1 \;,\label{5.7c}\end{eqnarray}
and the remaining ones are space-similar; for instance, $m_1^2- m_0^2=- (1+m_2^2+m_3^2)<0$.
    In the context of polarization optics, only this 4-vector is of interest and can be
    considered as representing the Stokes 4-vector.
    \begin{eqnarray}L\;  \left(\begin{array}{r} m_0\\m_1\\m_2\\m_3\end{array}  \right)=
        \left(\begin{array}{r} m_0\\-m_1\\-m_2\\-m_3\end{array}  \right) ;\label{5.8}\end{eqnarray}
and we may compare it with  (\ref{5.4d}).

Now, let us turn to general case of arbitrary Mueller matrices of Lorentz type
    \begin{eqnarray} L \; [ (A^*)^{-1}+  A^{-1} ]=A+A^* \;,\qquad
        L  \; [ (A^*)^{-1}-  A^{-1} ]= A-A^* \; ;\label{5.9a}\end{eqnarray}
Re-expressing the parameters $(q_0,q_1,q_2,q_3)$ in terms of their real and imaginary parts, we have
    \begin{eqnarray}q_0=x_0+i y_0 \;, \qquad   q_{j}=  x_{j}+i  y_{j}\;, \qquad
        q_0^*=x_0-i y_0 \;, \qquad   q_{j}^*=  x_{j} -i y_{j}\;,\nonumber\\
    x_0^2- x_1^2 -x_2^2- x_3^2-y_0^2+ y_1^2+ y_2^2+ y_3^2 =1,\qquad
        x_0y_0-x_1y_1-x_2y_2-x_3y_3 =0.\label{5.10}\end{eqnarray}
The relations (\ref{5.9a}) become equivalent to the following
    \begin{eqnarray}L \;\left(\begin{array}{rrrr} x_0&x_1&x_2&x_3\\x_1& x_0&-y_3&y_2\\
        x_2&y_3&x_0&-y_1\\x_3&-y_2&y_1&x_0\end{array}  \right)=
    \left(\begin{array}{rrrr} x_0&-x_1&-x_2&-x_3\\-x_1& x_0&y_3&-y_2\\
        -x_2&-y_3&x_0&y_1\\-x_3&y_2&-y_1&x_0\end{array}  \right) \;,\nonumber\\
    L \;\left(\begin{array}{rrrr} -y_0&-y_1&-y_2&-y_3\\-y_1& -y_0&-x_3&x_2\\
        -y_2&x_3&-y_0&-x_1\\-y_3&-x_2&x_1&- y_0\end{array}  \right)=
    \left(\begin{array}{rrrr} y_0&-y_1&-y_2&-y_3\\-y_1& y_0&-x_3&x_2\\
        -y_2&x_3&y_0&-x_1\\-y_3&-x_2&x_1&y_0\end{array}  \right) \; .\label{5.11b}\end{eqnarray}
Explicitly, the corresponding transitivity relations (\ref{5.9a}) read
{\small
    \begin{equation}\label{5.12b}\begin{array}{l}
    \psi_0=\left(\begin{array}{r} x_0\\x_1\\x_2\\x_3\end{array}  \right), \;
    L \psi_0=\psi'_0=\left(\begin{array}{r} x_0\\-x_1\\-x_2\\-x_3\end{array}  \right) \;,
    \psi_1=\left(\begin{array}{r} x_1\\  x_0\\  y_3\\ -y_2\end{array}  \right), \;
    L \psi_1=\psi'_1=\left(\begin{array}{rrrr}  -x_1\\   x_0\\  -y_3\\ y_2\end{array}  \right) \;,\medskip\\
    \Psi_2=\left(\begin{array}{r}   x_2\\ -y_3\\   x_0\\  y_1\end{array}  \right), \;
    L \Psi_2 =\Psi'_2 =\left(\begin{array}{rrrr}   -x_2\\ y_3\\   x_0\\  -y_1\end{array}  \right) \;,
    \Psi_3=\left(\begin{array}{r}  x_3\\  y_2\\ -y_1\\ x_0\end{array}  \right), \;
    L \Psi_3 =\Psi'_3 =\left(\begin{array}{r}  -x_3\\  -y_2\\ y_1\\  x_0\end{array}  \right) \;,\medskip\\
    L\Phi_0=\Phi'_0= \left(\begin{array}{r} y_0\\-y_1\\-y_2\\-y_3\end{array}  \right) \;,
    \Phi_1=\left(\begin{array}{r}  -y_1\\   -y_0\\ x_3\\ -x_2\end{array}  \right),
    L \Phi_1 =\Phi'_1= \left(\begin{array}{r}  -y_1\\  y_0\\ x_3\\ -x_2\end{array}  \right) \;,\medskip\\
    \Phi_2=\left(\begin{array}{r}  -y_2\\ -x_3\\   -y_0\\  x_1\end{array}  \right), \;
    L \Phi_2=\Phi'_2=\left(\begin{array}{r}   -y_2\\ -x_3\\  y_0\\  x_1\end{array}  \right) \;,
    \Phi_2=\left(\begin{array}{r}  -y_3\\  x_2\\ -x_1\\ -y_0\end{array}  \right), \;
    L \Phi_2=\Phi'_2 =\left(\begin{array}{r}  -y_3\\ x_2\\ -x_1\\   y_0\end{array}  \right) \; .
    \end{array}\end{equation}
}
It should be noted that the four cases describe transformations over 4-vectors
    which change the sign of the zero component into the inverse one, therefore
    the corresponding 4-vectors are space-similar ones and these cannot describe
    Stokes 4-vectors. Indeed, for an arbitrary time-similar vector, we have
    $$I. \qquad t^2 > x^2, \qquad t > x,\qquad t'={ e^{\beta}+e^{-\beta} \over 2 }
        t-{e^{\beta}-e^{-\beta} \over 2 } x > 0\; ;$$
where instead, for a space-similar vector, we  have
    \begin{eqnarray}II. \qquad t^2<x^2, \qquad \underline{t< x }, \qquad t'={ e^{\beta}+
        e^{-\beta} \over 2 } t-{e^{\beta}-e^{-\beta} \over 2 }x \qquad \Longrightarrow\nonumber\\
    \mbox{possible}\;\; t'<0,\qquad  \mbox{if}\qquad{ e^{\beta}+e^{-\beta} \over 2 } t<{e^{\beta}
        -e^{-\beta} \over 2 }x, \qquad{e^{2\beta}+1 \over e^{2\beta}-1}<{x \over t}\;.\nonumber\end{eqnarray}
As Stokes 4-vectors one can consider only the four vectors from the above: $\Psi_0, \Phi_1, \Phi_2, \Phi_3$.
{\bf Acknowedgements.} The present work was developed under the auspices of Grant
    1196/2012 - BRFFR - RA No. F12RA-002, within the cooperation framework between Romanian
    Academy and Belarusian Republican Foundation for Fundamental Research.

\vspace{10mm}

{\bf Elena Ovsiyuk and Olga  Veko}\\
Mosyr State Pedagogical University, Republic of Belarus.\\
E-mail: e.ovsiyuk@mail.ru , vekoolga@mail.ru,\\[1.6mm]

{\bf Mircea Neagu}\\
University Transilvania of Brasov, Romania.\\
E-mail: mirceaneagu73@gmail.com\\[1.6mm]

{\bf Vladimir Balan}\\
University Politehnica of Bucharest, Romania.\\
E-mail: vladimir.balan@upb.ro\\[1.6mm]

{\bf Victor Red'kov}\\
B.I. Stepanov Institute of Physics,\\
National Academy of Sciences of Belarus, Minsk, Republic of Belarus.\\
E-mail: v.redkov@dragon.bas-net.by
\end{document}